\documentstyle[epsf,12pt]{article}

\catcode`\@=11
\long\def\@makefntext#1{
\protect\noindent \hbox to 3.2pt {\hskip-.9pt
$^{{\ninerm\@thefnmark}}$\hfil}#1\hfill}		

\def\@makefnmark{\hbox to 0pt{$^{\@thefnmark}$\hss}}  

\def\ps@myheadings{\let\@mkboth\@gobbletwo
\def\@oddhead{\hbox{}
\rightmark\hfil\ninerm\thepage}
\def\@oddfoot{}\def\@evenhead{\ninerm\thepage\hfil
\leftmark\hbox{}}\def\@evenfoot{}
\def\sectionmark##1{}\def\subsectionmark##1{}}

\renewcommand{\thefootnote}{\fnsymbol{footnote}}

\newcounter{sectionc}\newcounter{subsectionc}\newcounter{subsubsectionc}
\renewcommand{\section}[1] {\vspace*{0.6cm}\addtocounter{sectionc}{1}
\setcounter{subsectionc}{0}\setcounter{subsubsectionc}{0}\noindent
	{\normalsize\bf\thesectionc. #1}\par\vspace*{0.4cm}}
\renewcommand{\subsection}[1] {\vspace*{0.6cm}\addtocounter{subsectionc}{1}
	\setcounter{subsubsectionc}{0}\noindent
	{\normalsize\it\thesectionc.\thesubsectionc. #1}\par\vspace*{0.4cm}}
\renewcommand{\subsubsection}[1]
{\vspace*{0.6cm}\addtocounter{subsubsectionc}{1}
	\noindent {\normalsize\rm\thesectionc.\thesubsectionc.\thesubsubsectionc.
	#1}\par\vspace*{0.4cm}}

\newcounter{appendixc}
\newcounter{subappendixc}[appendixc]
\newcounter{subsubappendixc}[subappendixc]

\renewcommand{\appendix}[1] {\vspace*{0.6cm}
        \refstepcounter{appendixc}
        \setcounter{figure}{0}
        \setcounter{table}{0}
        \setcounter{equation}{0}
        \renewcommand{\thefigure}{\Alph{appendixc}.\arabic{figure}}
        \renewcommand{\thetable}{\Alph{appendixc}.\arabic{table}}
        \renewcommand{\theappendixc}{\Alph{appendixc}}
        \renewcommand{\theequation}{\Alph{appendixc}.\arabic{equation}}
        \noindent{\bf Appendix \theappendixc #1}\par\vspace*{0.4cm}}

\def\abstracts#1{{

\centering{\begin{minipage}{12.2truecm}\footnotesize\baselineskip=12pt\noindent
	\centerline{\footnotesize ABSTRACT}\vspace*{0.3cm}
	\parindent=0pt #1
	\end{minipage}}\par}}

\renewenvironment{thebibliography}[1]
	{\begin{list}{\arabic{enumi}.}
	{\usecounter{enumi}\setlength{\parsep}{0pt}
\setlength{\leftmargin 1.25cm}{\rightmargin 0pt}
	 \setlength{\itemsep}{0pt} \settowidth
	{\labelwidth}{#1.}\sloppy}}{\end{list}}

\newcounter{itemlistc}
\newcounter{romanlistc}
\newcounter{alphlistc}
\newcounter{arabiclistc}

\newcommand{\fcaption}[1]{
        \refstepcounter{figure}
        \setbox\@tempboxa = \hbox{\footnotesize Fig.~\thefigure. #1}
        \ifdim \wd\@tempboxa > 6in
           {\begin{center}
        \parbox{6in}{\footnotesize\baselineskip=12pt Fig.~\thefigure. #1}
            \end{center}}
        \else
             {\begin{center}
             {\footnotesize Fig.~\thefigure. #1}
              \end{center}}
        \fi}

\newcommand{\tcaption}[1]{
        \refstepcounter{table}
        \setbox\@tempboxa = \hbox{\footnotesize Table~\thetable. #1}
        \ifdim \wd\@tempboxa > 6in
           {\begin{center}
        \parbox{6in}{\footnotesize\baselineskip=12pt Table~\thetable. #1}
            \end{center}}
        \else
             {\begin{center}
             {\footnotesize Table~\thetable. #1}
              \end{center}}
        \fi}

\def\@citex[#1]#2{\if@filesw\immediate\write\@auxout
	{\string\citation{#2}}\fi
\def\@citea{}\@cite{\@for\@citeb:=#2\do
	{\@citea\def\@citea{,}\@ifundefined
	{b@\@citeb}{{\bf ?}\@warning
	{Citation `\@citeb' on page \thepage \space undefined}}
	{\csname b@\@citeb\endcsname}}}{#1}}

\newif\if@cghi
\def\cite{\@cghitrue\@ifnextchar [{\@tempswatrue
	\@citex}{\@tempswafalse\@citex[]}}
\def\citelow{\@cghifalse\@ifnextchar [{\@tempswatrue
	\@citex}{\@tempswafalse\@citex[]}}
\def\@cite#1#2{{$\null^{#1}$\if@tempswa\typeout
	{IJCGA warning: optional citation argument
	ignored: `#2'} \fi}}

\def\@citeown#1#2{{${#1}$\if@tempswa , #2\fi }}

 1
 1
 1

\font\ninerm=cmr9

\begin{document}

\newcommand{\mycite}[1]{\raisebox{-1.85mm}{\large\cite{#1}}}

\centerline{\normalsize\bf MONOPOLES IN COMPACT U(1)}
\centerline{\normalsize\bf -- ANATOMY OF THE PHASE TRANSITION --}
\vspace*{0.6cm}
\centerline{\footnotesize A. Bode}
\baselineskip=13pt
\centerline{\footnotesize\it Physics Department, Humboldt University}
\baselineskip=12pt
\centerline{\footnotesize\it D-10115 Berlin, Germany}
\centerline{\footnotesize E-mail: achim@eiche.physik.hu-berlin.de}
\vspace*{0.3cm}
\centerline{\footnotesize K. Schilling\footnote{Talk presented by K.\
Schilling.}}
\baselineskip=13pt
\centerline{\footnotesize\it H\"ochstleistungsrechenzentrum HLRZ, KFA-
J\"ulich}
\baselineskip=12pt
\centerline{\footnotesize\it D-52425 J\"ulich, Germany}
\centerline{\footnotesize E-mail: schillin@wpts0.physik.uni-wuppertal.de}
\vspace*{0.3cm}
\centerline{\footnotesize V. Bornyakov}
\baselineskip=13pt
\centerline{\footnotesize\it Institute of High Energy Physics}
\baselineskip=12pt
\centerline{\footnotesize\it 142284 Protvino, Russia}
\vspace*{0.3cm}
\centerline{\footnotesize and}
\vspace*{0.3cm}
\centerline{\footnotesize Th. Lippert}
\baselineskip=13pt
\centerline{\footnotesize\it Physics Department, University of Wuppertal}
\baselineskip=12pt
\centerline{\footnotesize\it D-42097 Wuppertal, Germany}
\centerline{\footnotesize E-mail: lippert@wpts0.physik.uni-wuppertal.de}
\vspace*{0.9cm} \abstracts{We present evidence that the existence of
a first order phase transition in compact U(1) with Wilson action is not
related to
monopole loops wrapping around the toroidal lattice, as has been previously
suggested.  Our analysis is based on the suppression of such loops by `soft
boundary conditions' that correspond to an infinitely large chemical
potential for the monopoles on the boundary, during the updating process.
It is observed that the double peak structure characteristic for the first
order phase transition reappears at sufficiently large lattice sizes and
separations from the lattice boundary.}
\normalsize\baselineskip=15pt
\setcounter{footnote}{0}
\renewcommand{\thefootnote}{\alph{footnote}}

\section{Introduction}

During this workshop we have heard about plenty of indications that
confinement in SU(2) and SU(3) gauge theories can be traced back to an
underlying U(1) gauge structure. The dual Meissner effect mechanism
suggests monopoles to condense and confine the electric field into a narrow
flux tube.  It is widely believed---but still debated---that the maximal
abelian projection uncovers the relevant U(1) degrees of freedom in this
scenario.

In 4d compact U(1) lattice gauge theory with Wilson action, monopoles are
defined ab initio as gauge invariant objects, in contrast to non-abelian
theories.  It was demonstrated clearly in
ref.~\mycite{STACK} that large monopole loops
dominate the confining heavy quark potential at large distances in 4d
compact U(1) gauge theory.  This raises the question about the impact of
the dynamics of monopole loops onto the critical behaviour near the
confinement-deconfinement phase transition in U(1).

Despite the `simplicity' of the U(1) gauge theory, the order of its phase
transition---which is relevant to the existence of a continuum limit of the
lattice construction---has been under debate for 15 years by now, as we can
see from table~\ref{DEBATE}.
\begin{table}[htb]
\tcaption{Order of the U(1) phase transition.\label{DEBATE}}
\small
\begin{tabular}{||l|l|c|l|l||}
\hline\hline
author & max.\ lattice size & boundary conditions & year & order \\
\hline
Lautrup  et al.\cite{LAUTRUP}  & $H_4(6)$  & periodic & 1980 & $2^{nd}$ \\
M\"utter et al.\cite{MUTTER}   & $H_4(8)$  & periodic & 1982 & $2^{nd}$ \\
Jers\`ak et al.\cite{JERSAK}   & $H_4(8)$  & periodic & 1983 & $1^{st}$ \\
Bhanot   et al.\cite{GYAN}     & $H_4(16)$ & periodic & 1992 & $1^{st}$ \\
Lang     et al.\cite{LANG1,LANG2}& $SH_4(10)$& no boundary & 1993 &$2^{nd}$ \\
Baig     et al.\cite{BAIG}     & $H_4(16)$ & fixed    & 1994 & $2^{nd}$ \\
Lippert  et al.\cite{LIPPERT1} & $H_4(24)$ & ${\mbox{wrapping}\atop
                                               \mbox{loop suppressing}}$
                                                      & 1994  & $1^{st}$ \\
\hline\hline
\end{tabular}
\end{table}
Some years ago, it was argued in ref.~\mycite{JERSAK} that the lattice
heuristics of metastabilities is greatly affected by artifacts from the
periodicity on a finite lattice: the wrap-around of large monopole loops
(around the lattice with toroidal topology) might cause the go-slow of any
local updating algorithm and thus might fake first order behaviour on a
finite size system.  More recently evidence was presented in support of
this view by a study of a lattice with trivial homotopy group (without such
wrap-arounds!)\cite{LANG1,LANG2}. On a smallish lattice, equivalent in
volume to a periodic $10^4$ lattice, Lang and Neuhaus observed no sign of
first order behaviour.

In this contribution we intend to study the anatomy of this system in a
variant approach.  We will be less drastic in changing the lattice
topology, remaining on a 4d hypercubic lattice. We inhibit wrapping
monopole loops during the update by suppression of monopole currents to
traverse the boundary.  This is achieved by introduction of an infinitely
large chemical potential\cite{BORNYAKOV} on the boundary, that confines
monopole currents inside the volume, but allowing spin waves to cross.  The
work presented here is an extension of our previous studies\cite{LIPPERT1}
to a $32^4$ lattice.

\section{Monopole suppressing boundary conditions}

Following ref.~\mycite{DEGRAND},
we start out from the monopole current definition on the lattice,
\begin{equation}
  m_{\mu}(x)= \frac{1}{2}\epsilon_{\mu\nu\rho\sigma}\Big[
  n_{\rho\sigma}(x+\nu)-n_{\rho\sigma}(x)\Big],\qquad
  n_{\rho\sigma}=-2,-1,0,1,2.
\end{equation}
$n_{\mu\nu}(x)$ counts the number of Dirac sheets,
\begin{equation}
  n_{\mu\nu}(x)= \frac{1}{2\pi}\Big[ \bar\Phi_{\mu\nu}(x)-
  \Phi_{\mu\nu}(x)\Big],
\end{equation}
that is given by the difference between physical flux
$\bar\Phi_{\mu\nu}(x)$ and plaquette flux
$\Phi_{\mu\nu}(x)$.

As the updating procedure is sequential in terms of the links, we need to
clarify the interrelation between gauge links and monopole current
elements.  Let us consider the update of a link $U_{\rho}$ in a 3d cube
sitting on the boundary with normalvector $e_{\mu}$. We search for the
plaquettes associated with the Dirac sheets $n_{\rho\sigma}(x)$,
$n_{\rho\nu}(x)$ $n_{\rho,-\sigma}(x)$ and $n_{\rho,-\nu}(x)$ that
contribute to the monopole current $m_{\mu}(x)$ into this direction
$e_{\mu}$.  These plaquettes are depicted in Fig.~\ref{CUBE} with dotted
lines.
\begin{figure}
\hfill
\begin{minipage}[t]{5cm}
\epsfxsize=4 cm
\centerline{\epsfbox[33 587 236 773]{Figs/update.eps}}
\fcaption{Plaquettes contributing to $m_{\mu}(x)$.\label{CUBE}}
\end{minipage}
\hfill
\begin{minipage}[t]{5cm}
\epsfxsize=4 cm
\centerline{\epsfbox[95 209 561 576]{Figs/measure.eps}}
\fcaption{\sloppy Mea\-sure\-ments on nested shells (3d
projection).\label{NEST}}
\end{minipage}
\hfill\mbox{ }

\end{figure}
According to Fig.~\ref{CUBE} the update link $U_{\rho}$ belongs to
four 3d boundary cubes.

The monopole suppressing boundary condition is injected into the updating
procedure in the following way: if, for any of these cubes, the change in
flux is larger than $2\pi$, the proposed link change is rejected in the
Monte Carlo.  Therefore, the monopole current in $\mu$-direction in the
respective 3d cube is a conserved quantity in Monte Carlo time.  This
allows to enforce configurations with any given number of current elements
across the boundary. We initialize the simulation with {\em no} such crossings
at
all, which implies containment of monopole currents in the volume
throughout the simulation.

As the number of monopole currents in a boundary cube is a conserved
quantity the update rule presented fulfills {\em detailed balance}.  The
procedure of suppressing the monopoles can be viewed as being equivalent to
an infinitely large chemical potential for the monopoles on the cube where
the monopole current is suppressed: in the equivalent action
$
S = S_{W} + S_{\lambda},
$
$~S_{W}~$ represents the standard Wilson action and $~S_{\lambda}~$ is
the additional term which suppresses monopole currents $m_{\mu}$ at
the lattice slice $x_{\mu}=0$,
\begin{equation}
S_{\lambda} = \lambda \cdot \sum_{\mu} \sum_{x:~x_{\mu}=0} \mid
m_{\mu} (x) \mid , ~~\lambda \rightarrow \infty .
\end{equation}
Obviously, $S_{\lambda}$ explicitly breaks translation symmetry.

In Fig.~\ref{SUPP}, the monople content of a confined configuration taken
from a thermalized ensemble on an $8^4$ lattice at $\beta=1.0$ is plotted
for illustration.  Fig.~\ref{SUPP}a shows the situation on the standard
hypertorus and Fig.~\ref{SUPP}b refers to the implementation with monopole
containment.
\begin{figure}[p]
\epsfxsize=11 cm
\centerline{\epsfbox[83 5 504 788]{Figs/monos.eps}}
\fcaption{
a) Monopole cluster wrapping around the
   standard hyper-toroidal lattice and
b) suppression of currents across the boundaries leading
   to containment of the monopole clusters.
\label{SUPP}}
\end{figure}
In Fig.~\ref{SUPP}b, no monopole current can traverse the boundaries of the
lattice.  The required absence of traversing currents was monitored
throughout our simulation\footnote{In our 4d visualization, see
ref.~\mycite{BODEVIS}, the discrete euclidean time is represented by colors.
If an arrow traverses the spatial hyperplanes the colors of the arrow's
head and tip differ. The boundary in time direction is given by the color
pink.}.

\section{Results}

The simulation was carried out on a sequence of lattices, scanning for the
location of the phase transition, and in search for first order metastable
behaviour. The selected lattice sizes were $8^4$, $16^4$, $24^4$, and
$32^4$.  We performed about 100.000 sweeps at each $\beta$-value, that were
chosen within the suspected metastable region.  We used standard Metropolis
Monte Carlo updating.

We measured plaquettes and monopole density, with sampling performed on
nested shells (see the 3d projection of the shells in Fig.~\ref{NEST}), in
order to search genuine 4d bulk behaviour.  Notice that we have traded the
unwanted topological effects due to wrapping monopole loops against surface
effects, i.e. penetration phenomena! On a sufficiently large lattice, we
would expect with our  `experimental layout' to distinguish between
first and second order behaviour.

The results on the $8^4$ lattice showed no evidence whatsoever for first
order behaviour. On the $16^4$ lattice, we scan in $\beta$ through the
phase transition. As be seen from Fig.~\ref{plaq16}, the plaquette displays
little $\beta$-dependence in the outer shells.
\begin{figure}
\centerline{\epsfxsize=7.3cm\epsfbox[135 352 438
654]{Figs/plaq16.eps}\hfill\epsfxsize=7.3cm\epsfbox[135 352 438
654]{Figs/plaq24.eps}}
\begin{minipage}[t]{0.45\textwidth}
\fcaption{\sloppy Plaquette on shells of $16^4$-lattice as\\ function of
$\beta$.The square region
is drawn to\\
 indicate the size of Fig.~\ref{plaq24}.
\label{plaq16}}
\end{minipage}\hfill
\begin{minipage}[t]{0.45\textwidth}
\fcaption{Plaquette on shells of $24^4$-lattice as\\ function of $\beta$.
Note the change in scales\\
compared to Fig.~\ref{plaq16}.
\label{plaq24}}
\end{minipage}\hfill
\end{figure}
For the shells with distance greater than 4 from the boundary, however, we
discover a rapid transient behaviour. This indicates that the boundary
layer thickness close to phase transition is not prohibitively large for
our method to work. Detailed inspection (see Fig.~\ref{ser16}) of the time
series on the $8^4$ shell with best signal-to-noise ratio, however, does
not reveal indications for metastabilities.
\begin{figure}
\begin{minipage}[t]{0.48\textwidth}
\centerline{\epsfxsize=7.3cm\epsfbox[134 548 487 657]{Figs/ser16_6_0.990.eps}}
\centerline{\epsfxsize=7.3cm\epsfbox[134 548 487 657]{Figs/ser16_6_1.000.eps}}
\centerline{\epsfxsize=7.3cm\epsfbox[134 548 487 657]{Figs/ser16_6_1.002.eps}}
\centerline{\epsfxsize=7.3cm\epsfbox[134 548 487 657]{Figs/ser16_6_1.005.eps}}
\fcaption{\sloppy A $\beta$ scan of the time series  on the $8^4$ shell for the
                  $16^4$-lattice.
\label{ser16}}
\end{minipage}
\hfill
\begin{minipage}[t]{0.48\textwidth}
\centerline{\epsfxsize=7.3cm\epsfbox[134 548 487 657]{Figs/ser24_1_1.0060.eps}}
\centerline{\epsfxsize=7.3cm\epsfbox[134 548 487 657]{Figs/ser24_1_1.0065.eps}}
\centerline{\epsfxsize=7.3cm\epsfbox[134 548 487
657]{Figs/ser24_1_1.007125.eps}}
\centerline{\epsfxsize=7.3cm\epsfbox[134 548 487 657]{Figs/ser24_1_1.0080.eps}}
\fcaption{\sloppy $\beta$ scan of time series  on the outer $24^4$ shell for
the $24^4$-lattice.
\label{ser24_24}}
\end{minipage}
\end{figure}

We suspect that the $16^4$ lattice does not suffice to reveal true bulk
behaviour and proceed to a $24^4$ lattice.  The plaquette is plotted across
the phase transition in Fig.~\ref{plaq24}.  It shows a much steeper
behaviour on the inner shells. Note the change in scale in
Figs.~\ref{plaq16} and \ref{plaq24}. The region of Fig.~\ref{plaq24}
transforms to a square drawn in Fig.~\ref{plaq16}.

Turning now to the time history, we display the Monte Carlo dynamics as
seen on the $24^4$ boundary shell (see Fig.~\ref{ser24_24}), the $16^4$ shell
(see Fig.~\ref{ser24_16}), the $10^4$ shell (see Fig.~\ref{ser24_10}), and the
$4^4$ shell (see Fig.~\ref{ser24_4}), in a $\beta$ scan across the phase
transition.
\begin{figure}
\begin{minipage}[t]{0.48\textwidth}
\centerline{\epsfxsize=7.3cm\epsfbox[134 548 487 657]{Figs/ser24_5_1.0060.eps}}
\centerline{\epsfxsize=7.3cm\epsfbox[134 548 487 657]{Figs/ser24_5_1.0065.eps}}
\centerline{\epsfxsize=7.3cm\epsfbox[134 548 487
657]{Figs/ser24_5_1.007125.eps}}
\centerline{\epsfxsize=7.3cm\epsfbox[134 548 487 657]{Figs/ser24_5_1.0080.eps}}
\fcaption{\sloppy $\beta$ scan of time series  on the $16^4$ shell for
          the  $24^4$-lattice.
\label{ser24_16}}
\end{minipage}
\hfill
\begin{minipage}[t]{0.48\textwidth}
\centerline{\epsfxsize=7.3cm\epsfbox[134 548 487 657]{Figs/ser24_8_1.0060.eps}}
\centerline{\epsfxsize=7.3cm\epsfbox[134 548 487 657]{Figs/ser24_8_1.0065.eps}}
\centerline{\epsfxsize=7.3cm\epsfbox[134 548 487
657]{Figs/ser24_8_1.007125.eps}}
\centerline{\epsfxsize=7.3cm\epsfbox[134 548 487 657]{Figs/ser24_8_1.0080.eps}}
\fcaption{\sloppy $\beta$ scan of time series  on the $10^4$ shell for
          the $24^4$-lattice.
\label{ser24_10}}
\end{minipage}
\end{figure}
 From these figures we conclude that $\beta = 1.007125$ is the
most promising setting to study the critical dynamics.

Therefore let us take a closer look at $\beta=1.007125$.
Fig.~\ref{ser24_1.007125} presents the time series of the plaquette from
the $24^4$ down to the $4^4$ shell.
\begin{figure}
\begin{minipage}[t]{0.48\textwidth}
\centerline{\epsfxsize=7.3cm\epsfbox[134 548 487
657]{Figs/ser24_11_1.0060.eps}}
\centerline{\epsfxsize=7.3cm\epsfbox[134 548 487
657]{Figs/ser24_11_1.0065.eps}}
\centerline{\epsfxsize=7.3cm\epsfbox[134 548 487
657]{Figs/ser24_11_1.007125.eps}}
\centerline{\epsfxsize=7.3cm\epsfbox[134 548 487
657]{Figs/ser24_11_1.0080.eps}}
\fcaption{\sloppy $\beta$ scan of time series  on the $4^4$ shell for
          the $24^4$-lattice.
\label{ser24_4}}
\end{minipage}
\hfill
\begin{minipage}[t]{0.48\textwidth}
\centerline{\epsfxsize=7.3cm\epsfbox[134 548 487
657]{Figs/ser24_5_1.007125_f.eps}}
\centerline{\epsfxsize=7.3cm\epsfbox[134 548 487
657]{Figs/ser24_7_1.007125_f.eps}}
\centerline{\epsfxsize=7.3cm\epsfbox[134 548 487
657]{Figs/ser24_9_1.007125_f.eps}}
\centerline{\epsfxsize=7.3cm\epsfbox[134 548 487
657]{Figs/ser24_11_1.007125_f.eps}}
\fcaption{\sloppy Scan through different shells of plaquette time series  for
          the  $24^4$-lattice at $\beta=1.007125$.
\label{ser24_1.007125}}
\end{minipage}
\end{figure}
The fluctuations increase as one proceeds to the inner shells, due to a
decrease in self averaging. On top of those fluctuations, however, the
$8^4$ shell {\it clearly exhibits long metastabilities} of typical length
of 10000 Monte Carlo sweeps. On the $4^4$ shell, the fluctuations become so
large that the two state metastability signal starts to fade out.

To further improve the situation we extend our simulation to the $32^4$
lattice. The resulting pattern of time histories is similar.  But the
competing short range fluctuations are reduced and leave the long range
flip-flop behaviour more pronounced.

In order to reduce the high frequency modes in the time series, we apply a
simple filter technique\cite{NEUHAUS}, averaging the series over a sliding
window with length $l$, according to
$\tilde{P}(t_i)=1/l\sum_{j=1}^{l}P_{i+j-1}$.  The resulting histogram is
shown in Fig.~\ref{window}.  It provides definite evidence of a two-peak
structure characteristic for a first order phase transition.
\begin{figure}
\centerline{\epsfxsize=8cm\epsfbox[134 479 474 660]{Figs/window.eps}}
\fcaption{\sloppy Histogram of plaquette after smoothing of the
series on the  $32^4$-lattice at $\beta=1.0090$ for the $12^4$ shell.
\label{window}}
\end{figure}

\end{document}